\documentclass[lettersize,letter,twocolumn]{IEEEtran}
\usepackage{amsmath,amsfonts}
\usepackage{algorithmic}
\usepackage{algorithm}
\usepackage{array}
\usepackage[caption=false,font=normalsize,labelfont=sf,textfont=sf]{subfig}
\usepackage{textcomp}
\usepackage{stfloats}
\usepackage{url}
\usepackage{verbatim}
\usepackage{graphicx}
\usepackage{cite}

\usepackage{float}
\usepackage{pgfplots}
\usepackage{ctable}

\usepackage{caption}
\usepackage[justification=centering]{caption}

\newcommand{\bs}[1]{\ensuremath{\mathbf{#1}}}

\begin{document}

%\title{Improving the Thresholds of CC-GLDCPs}
\title{Improving the Thresholds of Generalized LDPC Codes with Convolutional Code Constraints}

\author{Muhammad Umar Farooq$^\dag$, Alexandre Graell i Amat$^\ddag$,~\IEEEmembership{Senior Member,~IEEE},\\ and Michael Lentmaier$^\dag$,~\IEEEmembership{Senior Member,~IEEE \vspace*{-5mm}}
\thanks{$\dag$Department of Electrical and Information Technology, Lund University, Lund, Sweden,
(emails:\{muhammad.umar\_farooq, michael.lentmaier\}@\\eit.lth.se),
%\vspace*{-0.5cm}
  $\ddag$Chalmers University of Technology, Gothenburg, Sweden. (email:alexandre.graell@chalmers.se).} \\%
\thanks{This work was supported in part by the Swedish Research Council (VR) under grant \#2017-04370. The simulations were performed on resources provided by the Swedish National Infrastructure for Computing (SNIC) at center for scientific and technical computing at Lund University (LUNARC).}}

% \author{
% \IEEEauthorblockN{Muhammad Umar Farooq$^\dag$, Alexandre Graell i Amat$^\ddag$,~\IEEEmembership{Senior Member,~IEEE},\\ and Michael Lentmaier$^\dag$, ~\IEEEmembership{Senior Member,~IEEE \vspace*{0mm}} \\
% \{muhammad.umar\_farooq,michael.lentmaier\}@eit.lth.se, alexandre.graell@chalmers.se}\\
% }

%\thanks{This paper was produced by the IEEE Publication Technology Group. They are in Piscataway, NJ.}% <-this % stops a space
%\thanks{Manuscript received April 19, 2021; revised August 16, 2021.}}

\markboth{Submitted to IEEE Communications Letters, November~2022}{}


\maketitle

\begin{abstract}
CC-GLPDC codes are a class of generalized low-density parity-check (GLDPC) codes where the constraint nodes (CNs) represent convolutional codes. This allows for efficient decoding in the trellis with the forward-backward algorithm, and the strength of the component codes easily can be  controlled  by the encoder memory without changing the graph structure.

In this letter, we extend the class of CC-GLDPC codes by introducing different types of irregularity at the CNs and investigating their effect on the BP and MAP decoding thresholds for the binary erasure channel (BEC). 
For the considered class of codes, an exhaustive grid search is performed to find the BP-optimized and MAP-optimized ensembles and compare their thresholds with the regular ensemble of the same design rate. The results show that 
 irregularity can significantly improve the BP thresholds, whereas the  thresholds of the MAP-optimized ensembles are only slightly different from the regular ensembles. Simulation results for the AWGN channel are presented as well and compared to the corresponding thresholds. 

%Results show that for low-rate CC-GLDPC codes, a trade-off between the BP and the MAP optimized CC-GLDPC codes offer a moderate BP decoding performance, with steep slopes in the waterfall region and no visible error-floor.
%For high-rate CC-GLDPC codes, the MAP optimized CC-GLDPC codes seem to be the preferred design choice. MAP optimized CC-GLDPC codes are in general observed to have structure matching to a regular graph, whereas BP optimized CC-GLDPC codes are composed of an irregular graph.
\end{abstract}


% fix design rate, introduce irregularity at the check nodes

\section{Introduction}
%\IEEEPARstart{T}{urbo}~\cite{berrou_turbocodes}

Low-density parity-check codes (LDPC) codes are known for their excellent iterative decoding performance at large block lengths.  Generalized LDPC (GLDPC) codes are obtained by replacing the weak single parity-check codes at the constraint nodes (CNs) by more powerful component codes. It has been demonstrated in \cite{Yanfang_David_GLDPC_urllc} that this can be of advantage in the short block length regime, where LDPC codes suffer from a larger number of short cycles in their Tanner graph. Asymptotically, GLDPC codes based on  regular graphs have very good minimum distance growth rates and  MAP decoding thresholds, even if the variable node (VN) degree is only two. These properties make them suitable for spatial coupling \cite{mitchellSCGLDPC2021a} to benefit from improved BP decoding thresholds due to the threshold saturation effect. Also staircase codes \cite{smithStaircaseCodesFEC2012}, designed for high-speed optical communications, can be viewed as a class of  spatially coupled  GLDPC codes with higher density and product code structure, similar to braided codes \cite{zhangSpatiallyCoupledSplitComponent2018a}.

Without spatial coupling, analogously to LDPC codes, %~\cite{RichardsonIrregularLDPC}, 
the BP decoding thresholds of GLDPC codes can be improved with irregular graphs, either on the VN or the CN side~\cite{paoliniGeneralizedDoublyGeneralized2010}.  A simple example of check-hybrid GLDPC codes \cite{mulhollandMinimumDistanceDistribution2013} is the code doping procedure introduced in \cite{livaQuasicyclicGeneralizedLdpc2008}, where some CNs in the protograph of a given LDPC code are replaced by stronger CNs. This procedure is extended in \cite{Yanfang_David_GLDPC_urllc} to replacing a tunable fraction of CNs for code optimization. 
A common challenge in the above mentioned designs of GLDPC codes is that the design rate of the ensembles is changing when CNs are replaced, which makes a fair comparison between  codes difficult. Furthermore, it is not straightforward to increase the strength of a component code without changing its rate and length, which in turn would affect the graph structure.   

In \cite{muf_gldpctrellis} we introduced a family of regular GLDPC codes with convolutional code (CC) constraints that avoids this challenge.  These CC-GLDPC codes can be viewed as turbo-like codes with regular graph structure \cite{Moloudi_SCTC_Journal} and hence as an extension of braided convolutional codes to arbitrary regular graphs.
The aim was to identify the performance trade-off between CC-GLDPC codes with strong component codes and lower VN degree to conventional LDPC codes with larger VN degree.

It was observed that both the minimum distance of CC-GLDPC codes and their MAP thresholds improve with either the strength of the component code trellis or with increasing the VN degree of the graph at the expense of degraded BP thresholds.  
This type of performance behavior make CC-GLDPC codes suitable for spatial coupling.
Indeed, the SC-CC-GLDPC codes were observed numerically to exhibit the threshold saturation phenomenon.

In this letter, we introduce ensembles of irregular CC-GLDPC codes where the CNs represent component codes that may differ in memory, rate and fraction of edges they are connected to. Based on the observations in \cite{muf_gldpctrellis}, we intentionally keep the VN degree low but avoid degree-1 VNs which are typical for classical turbo-like codes \cite{Moloudi_SCTC_Journal}. More precisely, we use regular degree-two VNs and limit the number of CN types to two. Considering the design rates $R=1/3, 1/2,$ and $2/3$ as example, we perform a density evolution analysis on the binary erasure channel (BEC) and search for the ensemble parameters that lead to the best BP and MAP thresholds, respectively. Finally,  bit-error-rate (BER) simulations are carried out to validate the thresholds.

%----------------------------- Section II: Background
\section{Ensembles of Irregular GLDPC Codes with Convolutional Code Constraints}

\begin{figure}[!t]
    \centering
    \includegraphics[width=0.5\textwidth]{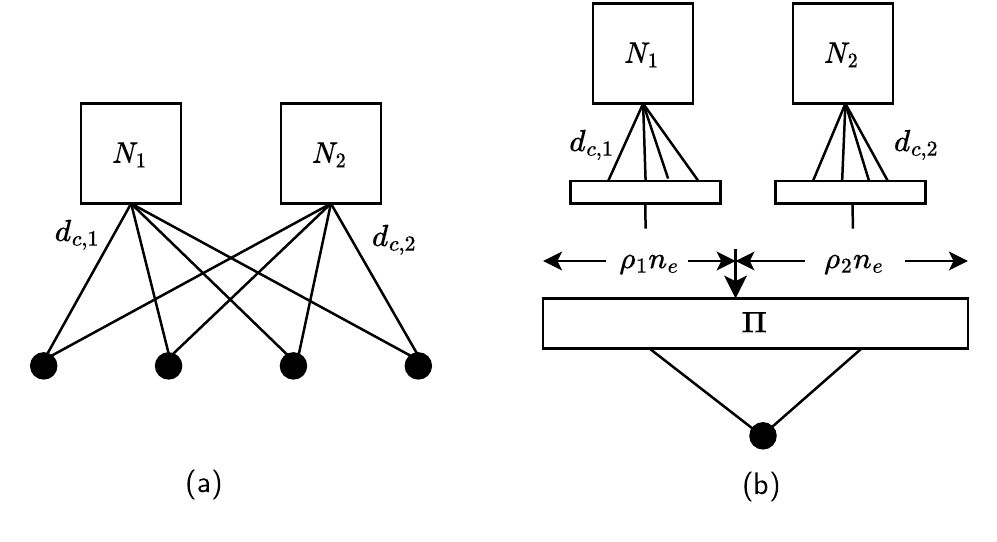}
    \caption{Compact graph: (a) (2,4) regular ensemble (b) proposed ensemble with adjustable CNs.}   \label{fig:compactgraphs}
\end{figure}

\subsection{Compact Graph Representation}

CC-GLDPC codes, analogously to protograph-based LDPC codes, can be described by means of a compact graph representation, as shown in Fig.~\ref{fig:compactgraphs}(a) for a $(2,4)$-regular ensemble \cite{muf_gldpctrellis}. In general, each VN corresponds to a block of code symbols and each CN to a convolutional  code $\mathcal{C}_j$ of rate $r_j=k_j/n_j$ and memory $m_j$, represented by a trellis of length $N_j$. The degree $d_{\text{c},j}$ of a CN is equal to the number of code symbols per trellis section, i.e., $d_{\text{c},j}=n_j$. 

\begin{figure}[!t]
    \centering
    \includegraphics[width=0.45\textwidth]{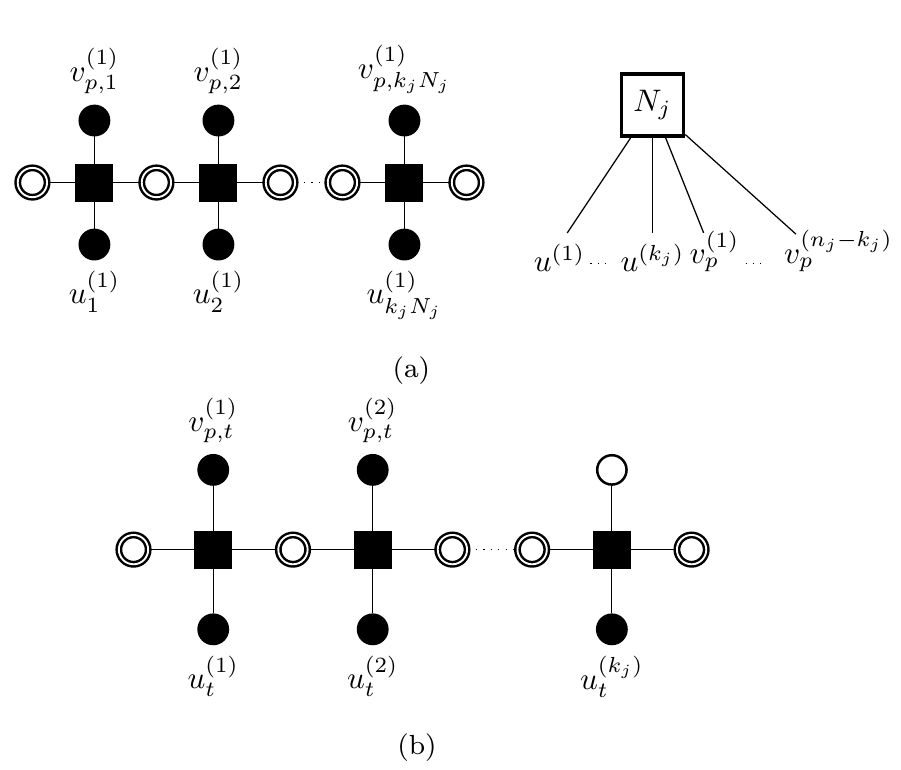}
    \caption{Factor graph representation of a CN trellis (a) Rate $1/2$ trellis (left) in a CN (right). (b) Trellis section at time $t$ after puncturing to rate $k_j/n_j$.}   \label{fig:puncturedtrellis15}
\end{figure}

The value $N_j$ is equivalent to the lifting factor of a protograph node, and the number of edges $n_{\text{e},j}=d_{\text{c},j} \cdot N_j$ in the lifted graph are equal to the total number of code symbols in the corresponding trellis. The permutations occur along the edges of the compact graph.  
A component code $\mathcal{C}_j$ of rate $r_j=k_j/n_j \geq 1/2$ can be constructed by puncturing a rate $r_{\text{m}}=1/2$ mother code trellis with $k_jN_j$ sections.
The factor graph of the mother code trellis is illustrated in Fig.~\ref{fig:puncturedtrellis15}(a), together with  the corresponding CN. The desired rate is achieved by combining $k_j$ sections to a rate $k_j/(2k_j)$ code and puncturing $2k_j-n_j$ of the parity bits. 
The punctured bits are shown as non-shaded VNs in Fig.~\ref{fig:puncturedtrellis15}(b).

Within this paper we assume that the number of CNs is equal to two and all VNs have degree $d_{\text{v}}=2$, i.e., both information and parity symbols are protected twice by a trellis. The resulting overall codes have length $n=(n_{\text{e},1}+n_{\text{e},2})/2$. In order to improve the decoding thresholds we consider different types of irregularity at the CN side while keeping the VNs regular.

\subsection{Irregular Component Code Memory}

Under a preserved graph structure, in \cite{muf_gldpctrellis} the strength of the code ensemble was changed by varying the component code strength.
Component code trellises with generator polynomials having trellises memories of one, two and three, respectively were used.
%The ability to alter the strength of the component codes without changing the graph structure is a distinguishing feature of trellis based GLDPC codes that separates them from the block component codes based GLDPC codes.

% To define the concept of irregularity in the component code strength, we first consider a regular graph structure, with $N_1 = N_2$, and $d_c^{(1)} = d_c^{(2)}$ in Figure~\ref{fig:g_irregular}.
 
An irregular CC-GLDPC w.r.t. the component code trellis strength is obtained when $m_1 \ne m_2$, whereas a regular CC-GLDPC w.r.t. the component code trellis strength is obtained when $m_1=m_2$. 
In this work, we have utilized the trellises with memories $m_j=1,2,3$ having generator polynomials $(1,1/3)$,  $(1,5/7)$, and $(1,13/15)$.
% As an example, the trellis strengths pairs of $\bs{m} = (m^{(1)},m^{(2)}) =\{(1,1),(2,2),(3,3)\}$ are regular, whereas $\bs{m} = \{(1,2), (1,3) (2,3), (2,1), (3,1), (3,2) \}$ are irregular.
An advantage of trellis-based GLDPC codes is that the strength of the component codes can be changed without altering the graph structure, or equivalently the design rate $R$ of the graph.

\subsection{Irregular Distribution of Edges}
% now proportions change, N1 different from N2

Now consider the case that $d_{\text{c}} = d_{\text{c},1}= d_{\text{c},2}$, $N_1 \ne N_2$, and $m_1 \ne m_2$.
In this case, both the strengths and the proportions of the  component codes $\mathcal{C}_j$ become irregular, but their rates are kept regular. 
Let us introduce $\rho_j$, which represents the fraction of the total number of edges that are connected to $\mathcal{C}_j$,
which can be  expressed as
\begin{equation}
    \rho_1 = \rho = \frac{d_{\text{c},1} \cdot N_1 }{d_{\text{c},1} \cdot N_1 + d_{\text{c},2} \cdot N_2 } \ , \quad \rho_2 = 1- \rho  \ .
\end{equation}

A varying $\rho$ produces a family of codes with a varying proportion of $\mathcal{C}_j$, under a fixed $R = 1 - 2/d_{\text{c}}$, and fixed component code rates $r= r_1 = r_2$.  

\subsection{Irregular Component Code Rates}

For a given design rate $R$, an irregularity w.r.t. the component code rates pair $\bs{r} = (r_1,r_2)$ is achieved if $r_1 \ne r_2$, whereas regularity w.r.t. the rates is achieved if $r_1 = r_2$.
Regular or irregular component code rates $r_j$ can be combined with regular or irregular memories $m_j$, and  regular or irregular lifting factors $N_j$ under the constraint that the design rate $R$ remains fixed.
The design rate in terms of component code rates pair $\bs{r} = (r_1,r_2)$, is expressed as
\begin{equation} \label{eq:crategen}
\begin{split}
        R(\rho,\bs{r}) &= 1-d_v(1-r_c(\rho,\bs{r})) \\
                        &= 2r_c(\rho,\bs{r}) - 1  \ ,
\end{split}
\end{equation}
where  $r_c$ denotes the average component code rate
\begin{equation} \label{eq:averagerc}
    r_c(\rho, \bs{r}) = \rho \, r_1 + (1-\rho) \, r_2 \ .
\end{equation}
For a given $r_1$ and a fixed design rate $R$ we can then express $r_2$ as a function of $\rho$, namely
\begin{equation} \label{eq:r2_rho}
    r_2(\rho) = \frac{r_c-\rho \cdot r_1}{(1-\rho)} =  \frac{(R+1)/2 -\rho \cdot r_1}{(1-\rho)}\ .
\end{equation}

We remark that both $r_1$ and $r_2$ are rational fractions, and lie within the interval $[r_m,1)$, where $r_m = 1/2$ is the rate of the mother code trellis used in this work.
For a given design rate $R$ or an averaged component code rate $r_c$, a family of CC-GLDPC codes with irregular component codes is obtained for $\rho \in [0,1]$.

%------------------------------------------------------------

\section{Threshold Analysis and Optimization} \label{sec:caccgldpc}
% Irregular LDPCs are shown to achieve capacity in the BP decoding, in which the optimized degree distributions are searched by employing the stochastic or linear optimization techniques.
% Irregular CC-GLDPC code ensemble may potentially be well suited to achieve capacity in the BP decoding sense.

In this section, we perform a decoding threshold analysis and perform an exhaustive grid search to determine the ensemble parameters resulting in the most competitive BP and MAP thresholds. Such a full search becomes feasible for the BEC, where exact density evolution equations can be derived. For the optimized ensembles, AWGN channel thresholds will then be given in Section~\ref{sec:finitelength} and compared to BER simulations.

% To make the exhaustive search feasible, the component code rate dimension in the search space is discretized. 
% The granularity or resolution of the search grid is varied by varying the individual lifts.

\subsection{Density Evolution Equations}
The mother code transfer functions of the considered generator polynomials can be derived by following the method explained in~\cite{Moloudi_SCTC_Journal}.
Suppose $f_{\text{s}}$ and $f_{\text{p}}$ denote the decoder transfer functions of systematic and parity bits of the mother code, then
\begin{equation}\label{eq:transfermother}
    p_{\text{s}}^{(i)} = f_{\text{s}}(q_{\text{s}}^{(i)},q_{\text{p}}^{(i)}),
\end{equation}
\begin{equation}
    p_{\text{p}}^{(i)} = f_{\text{p}}(q_{\text{s}}^{(i)},q_{\text{p}}^{(i)}),
\end{equation}
where $p_{\text{s}}^{(i)}$ and  $p_{\text{p}}^{(i)}$ are the erasure probabilities of the outgoing extrinsic message from the CN, and the $q_{\text{s}}^{(i)}$ and $q_{\text{p}}^{(i)}$ are the erasure probabilities of the incoming extrinsic messages to the CN.
If the parity bits of the mother code trellis are punctured randomly with a uniform distribution to achieve the desired component code rate, then the incoming erasure probabilities to the transfer functions become
\begin{equation}
    q_{\text{s}}^{(i)} = q^{(i-1)},
\end{equation}
\begin{equation}
    q_{\text{p}}^{(i)} = \frac{2k_j-n_j}{k_j} \cdot 1 + \frac{n_j-k_j}{k_j}  q^{(i-1)} \ .
\end{equation}
After the decoding, the erasure probability of the averaged extrinsic message from $\mathcal{C}_j$  is computed using
\begin{equation} \label{eq:transferFuncPunc}
   p_j^{(i)} = \frac{k_j}{n_j} \cdot  p_{\text{s}}^{(i)} + \frac{n_j - k_j}{n_j} \cdot  p_{\text{p}}^{(i)}\ .
\end{equation}
Combining the contributions of both decoders we get
\begin{equation} 
 p^{(i)} = \rho_1 p_1^{(i)}  + \rho_2 p_2^{(i)} = f_{\text{T}}(q^{(i-1)}) \ , 
\end{equation}
where the transfer function $f_{\text{T}}(\cdot)$ follows from applying  equations \eqref{eq:transfermother}--\eqref{eq:transferFuncPunc}.
Finally, including the VN update the DE recursion can be expressed as
\begin{equation}
    q^{(i)} = \varepsilon \cdot f_{\text{T}}(q^{(i-1)}) \ .
\end{equation}

The transfer function of the punctured trellis in~\eqref{eq:transferFuncPunc} is applicable for a rate range of $\left(1/n^{(j)},\cdots, (n^{(j)}-1)/n^{(j)} \right)$.

\subsection{Computation of BP and MAP Thresholds}

We use the BP-EXIT function $h^{\text{BP}}(\varepsilon)$~\cite{RU_mct} to determine the BP and MAP thresholds of CC-GLDPC codes, which in parameterized form is expressed as
\begin{equation*}
   h^{\text{BP}}(\varepsilon) = 
\left\{
	\begin{array}{ll}
		(\varepsilon,0) & \varepsilon \in [0,\varepsilon^{\text{BP}}) \\
		(\varepsilon(x),f_{\text{T}}^2(x)  & \varepsilon \in (\varepsilon^{\text{BP}},1] \leftrightarrow x \in (x^{\text{BP}},1]
	\end{array}
\right.
\end{equation*}
since we restrict the VN degree to   $d_\text{v}=2$ in this work.
Note that $x^{\text{BP}}$ corresponds to value of $x$ for which $\varepsilon(x)$ is minimum.

The MAP threshold is determined from the BP-EXIT function as
\begin{equation*}
    R(\rho,\bs{r}) = \int_{\varepsilon^{\text{MAP}}}^{1} h^{\text{BP}}(\varepsilon) \, d\varepsilon \ .
\end{equation*}

\subsection{Threshold Optimization}

A  parameter search space for the design of CC-GLDPC codes is  given by the CN parameters: rate pair $\bs{r}$, trellis memory pair $\bs{m}$, %degrees pair $\bs{d}_c$ or equivalently the
and fraction pair $(\rho,1-\rho)$.
Hereby the possible choices of component code rates and the fraction $\rho$ for a given design rate $R$ are interlinked via $\eqref{eq:averagerc}$.
The restriction to a code graph with two CNs and degree-two VNs results in a symmetry in the component code rates.
This symmetry allows us to reduce the search space parameters to $r_1$, $\rho$, and $\bs{m}$.
Note that $r_2$ is computed from~\eqref{eq:r2_rho} for a given $\rho \in [0,1]$ and $r_1 \in [r_c,1)$. 
The graph symmetry in component code rate implies that for a fixed $\rho$, when $r_2$ lies within $[r_m,r_c]$, then the resultant $r_1$ lies within $[r_c,1)$.
This holds even when the rate ranges described above are swapped.

Along the component code rate pair $\bs{r}=(r_1,r_2)$ and the fractions pair $(\rho,1-\rho)$ the parameters are continuous, and to limit the search space we use discretization by considering a lifted code graph with a total number of edges $n$. Sweeping the parameter $\rho=2/n,\dots,(n-2)/n$ we then obtain a finite number of possible rate pairs $(r_1,r_2)$ that make an exhaustive grid search possible. The size of $n$ determines the resolution of the component code rates and fractions, and it has to be chosen carefully to balance accuracy and complexity. In our threshold optimization we use values of $n$ close to 512.

\subsection{Results and Discussion}

Thresholds of CC-GLDPC codes having regular graph structure $d_{\text{c},1}  = d_{\text{c},2} $, $N_1 = N_2$, and both regular and irregular trellis strengths are shown in Table~\ref{table:c1family}\,(left).
It can be observed that irregularity w.r.t. trellis strength alone, while keeping the component codes degrees, and rates regular, doesn't offer performance improvement compared to the fully regular CC-GLDPC codes.

\begin{figure}[t]
    \centering
    \includegraphics[width=0.5\textwidth]{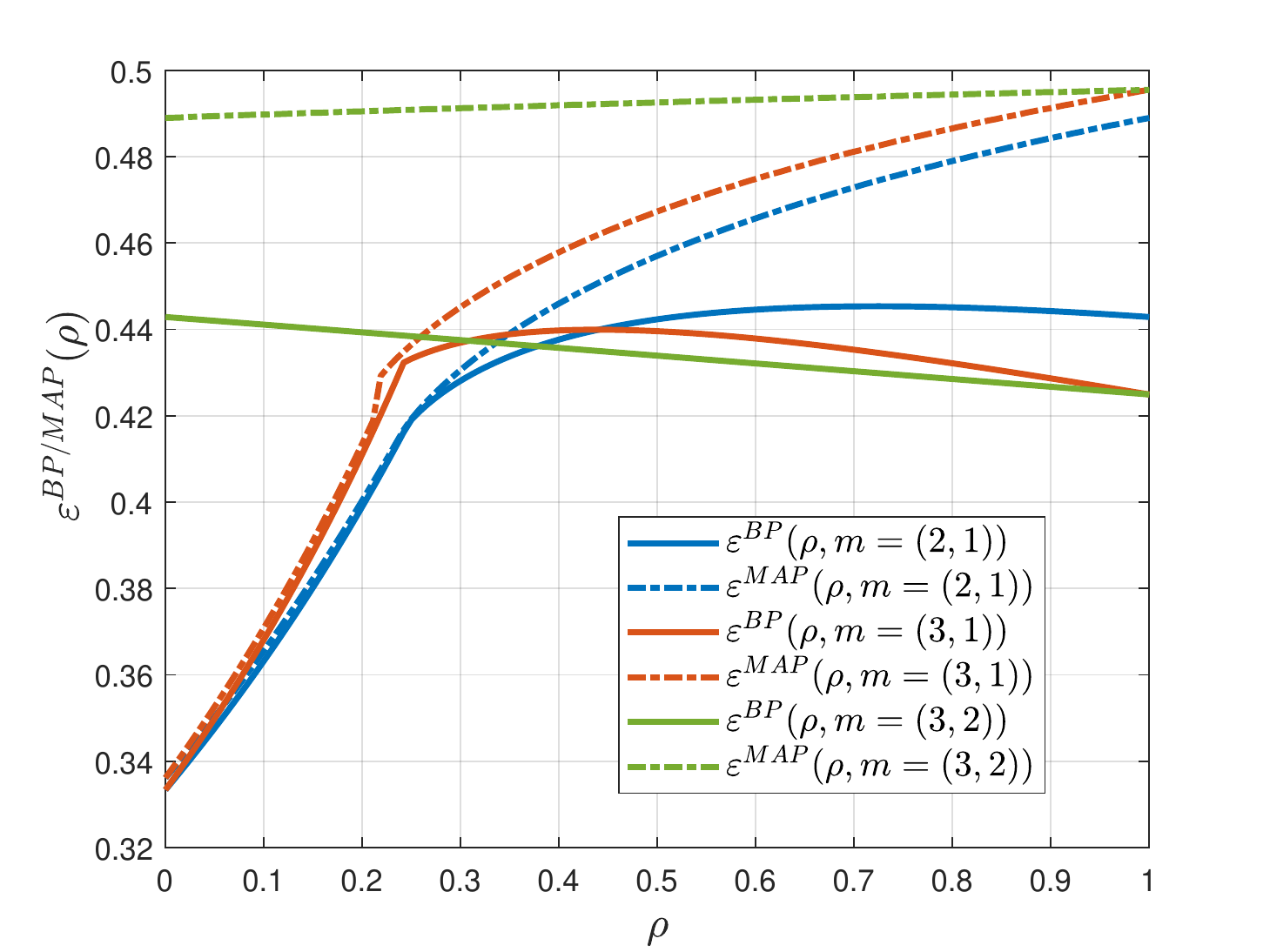}
    \caption{BP and MAP thresholds vs $\rho$ of rate $R=1/2$ CC-GLDPC codes with irregular trellis strength.}
    \label{fig:thdsVfractions}
\end{figure}

 Figure~\ref{fig:thdsVfractions} shows how an irregular distribution of edges, characterized by the parameter $\rho$, influences the  thresholds. Choosing the best parameter $\rho=\rho^*$ results in the values shown in Table~\ref{table:c1family}\,(right). In this case the BP threshold can be improved compared to the fully regular case.

\begin{table}[!t] 
\caption{Thresholds of rate $R=1/2$ CC-GLDPC codes with irregular trellis strength.}
\begin{center}
\begin{tabular}{cll|ll}
\toprule
$(m_1,m_2)$,  & $\varepsilon^{BP}$ & $\varepsilon^{MAP}$  & $\rho^*$ &  $\varepsilon^{BP}$ \\
\midrule
$(1,1)$         &    0.3334        & 0.3361  & -    & -\\
$(2,1)$     &    0.4423       &      0.4570   &  $103/128$   & 0.4453 \\
$(3,1)$        &    0.4396        &     0.4673   & $56/128$  & 0.4399 \\
$(2,2)$         &    0.4429        &  0.4889    & -  & -\\
$(3,2)$       &    0.4339        &     0.4926   & $0$  & 0.4429\\
$(3,3)$        &    0.4249        &     0.4955    & - & - \\
\bottomrule
\end{tabular}
\label{table:c1family}
\end{center}
\end{table}

Now we compute BP and MAP thresholds of rate $R =1/3,1/2,2/3$ CC-GLDPC codes over the complete search space, including irregular component code rates.  
For a given trellis memory pair \bs{m}, the exhaustive search produces a family of curves $\varepsilon^{\text{BP}}$ as a function of $r_1(\rho)$ for each $\rho$.
The design configuration with the best BP decoding performance then corresponds to the largest BP threshold in the family of curves $\varepsilon^{\text{BP}}(r_1(\rho,\bs{m}))$.

\begin{figure}[!t]
    \centering
    \includegraphics[width=0.5\textwidth]{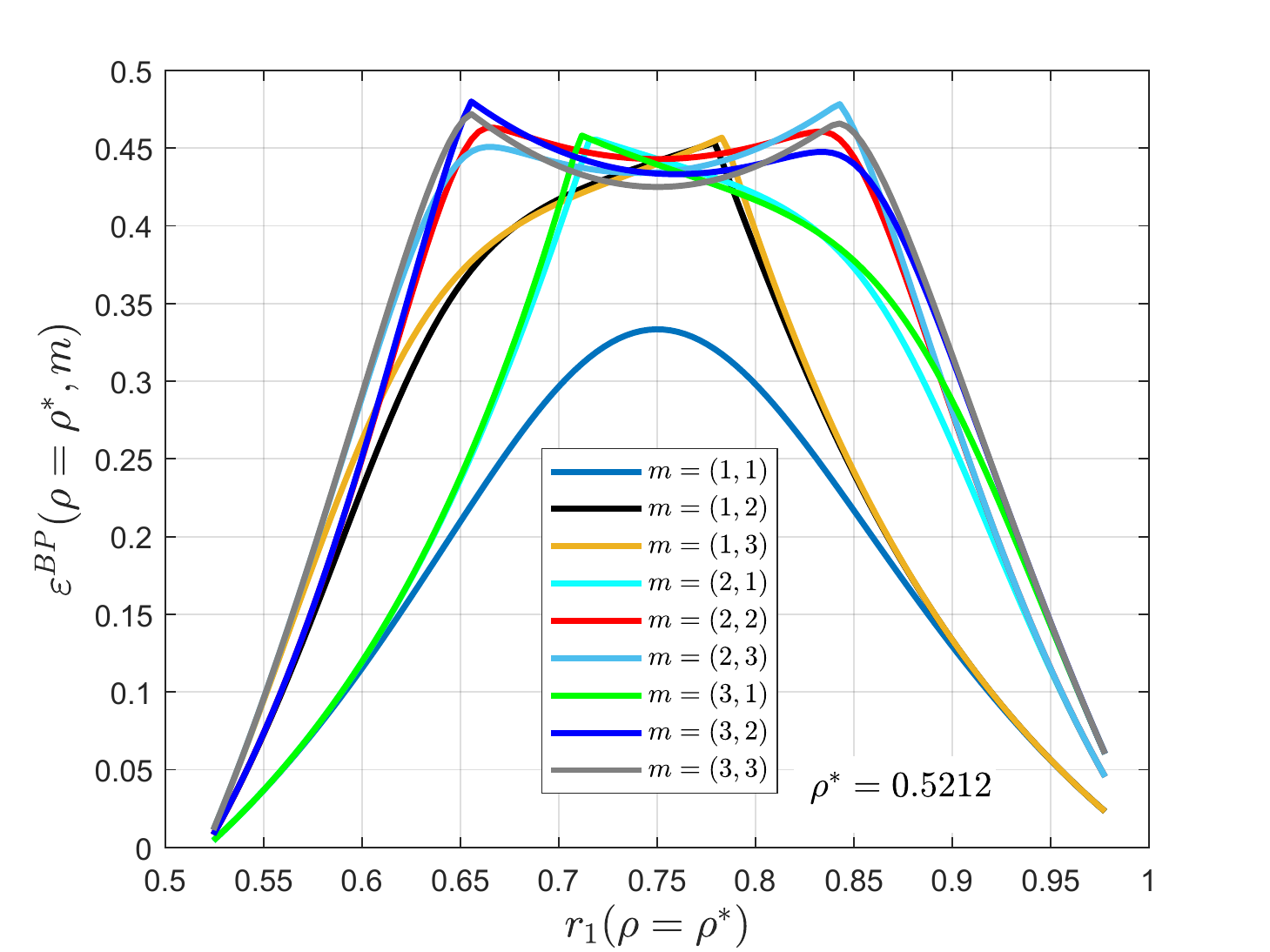}
    \caption{BP Threshold vs rate at $\rho^*$ of rate $R=1/2$ CC-GLDPC codes with irregular component code rates.}
    \label{fig:rateVrho_BPmax}
\end{figure}

Suppose $\rho^*$ denotes the  fraction that produces the best BP threshold in the considered parameter search space.
The plot of $\varepsilon^{\text{BP}}$ as a function of $r_1(\rho=\rho^*,\bs{m})$ for $R = 1/2$ CC-GLDPC is shown in Fig.~\ref{fig:rateVrho_BPmax}.
The plot is obtained by projecting the curves $\varepsilon^{\text{BP}}$ of all trellis memory combinations into a single dimension.
It can be seen from the figure that $\rho^* = 0.5212$, $\bs{m}=(3,2)$, and $r_1 = 0.6554$ constitute the design parameters corresponding to the best BP threshold of rate $R=1/2$ CC-GLDPC codes.
Similarly, we obtain the design configuration with best MAP decoding performance.

The design parameters corresponding to best BP and best MAP thresholds for $R=1/3, 1/2,$ and $2/3$ are listed in Table~\ref{table:r131223bestBPMAP}.
It is observed that the irregular CC-GLDPC graph structures yield the best BP and MAP decoding thresholds.
However, the MAP decoding performance of the regular code ensembles is very similar to that of the MAP-optimized irregular ones, also visible in Table~\ref{table:r131223bestBPMAP}.

\begin{table}[!t] 
\caption{Design Parameters of Best Thresholds}
\begin{center}
\begin{tabular}{llll}
\toprule
 Parameter &   Best BP & Regular & Best MAP \\
\midrule
\multicolumn{4}{c}{Rate-$1/3$ CC-GLDPC}\\
\midrule
$\varepsilon^{BP}$  & 0.6441    & 0.5352 & 0.5356 \\
$\varepsilon^{MAP}$  & 0.6603   &  0.6659  & 0.6660  \\
$(\rho^*,1-\rho^*)$ & (0.4773,0.5227) &(0.5,0.5) &(0.3295,0.6705)\\
$(r_1,r_2)$   & (0.5159,0.8043) &(0.6667,0.6667) &(0.6552,0.6723)\\
$(m_1,m_2)$  & (2,2)  &(3,3) & (3,3)\\
\midrule
\multicolumn{4}{c}{Rate-$1/2$ CC-GLDPC}\\
\midrule
$\varepsilon^{BP}$  & 0.4799    &   0.4249 & 0.4250 \\
$\varepsilon^{MAP}$  & 0.4858   &   0.4955 & 0.4955 \\
$(\rho^*,1-\rho^*)$ & (0.5215,0.4785) & (0.5,0.5) & (0.2266,0.7734)\\
$(r_1,r_2)$   & (0.6554,0.8531) & (0.75,0.75) & (0.7414,0.7525)\\
$(m_1,m_2)$  & (3,2)  & (3,3) & (3,3)\\
\midrule
\multicolumn{4}{c}{Rate-$2/3$ CC-GLDPC}\\
\midrule
$\varepsilon^{BP}$  & 0.3041    &  0.2893 & 0.2893\\
$\varepsilon^{MAP}$  & 0.3093   &  0.3189 & 0.3189\\
$(\rho^*,1-\rho^*)$ & (0.8274,0.1726) & (0.5,0.5) & (0.1726,0.8274)\\
$(r_1,r_2)$   & (0.8082,0.9540) & (0.8333,0.8333) &(0.8276,0.8345)\\
$(m_1,m_2)$  & (3,3)  & (3,3) & (3,3)\\
\bottomrule
\end{tabular}
\label{table:r131223bestBPMAP}
\end{center}
\end{table}

%------------------  section: Performance Analysis
\section{Finite Length Performance}\label{sec:finitelength}
We have performed BER simulations for CC-GLDPC codes of length $n=102400$  with the ensemble parameters corresponding to the best BEC thresholds w.r.t. the BP and MAP decoding performance listed in Table~\ref{table:r131223bestBPMAP}.
%Some  strategies leading to a trade-off between the BP thresholds and error floors are also discussed.  
The simulations are compared with AWGN channel thresholds.
 Table~\ref{table:r131223predictedthd} shows AWGN channel thresholds computed by the erasure channel prediction  method used in~\cite{muf_thdbccawgnc}, while the thresholds in Table~\ref{table:r13awgnthd} are obtained by Monte Carlo DE based on histograms.

\begin{table}[!t] 
\caption{Predicted AWGN Thresholds $E_b/N_0(\rho^{*}) \, [\mathrm{dB}]$}
\begin{center}
\begin{tabular}{llrrr}
\toprule
$R$ & Parameter &   Best BP & Regular & Best MAP \\
\midrule
$1/3$ & $\varepsilon^{BP}$  & -0.1261    & 1.4747 & 1.4692 \\
$1/3$ & $\varepsilon^{MAP}$  & -0.3893   &-0.4824  & -0.4841  \\
\midrule
$1/2$ & $\varepsilon^{BP}$  & 0.4511   &  1.1565  & 1.1548 \\
$1/2$ & $\varepsilon^{MAP}$  & 0.3740  &  0.2467  &  0.2467 \\
\midrule
$2/3$ & $\varepsilon^{BP}$ & 1.4313  &  1.6216  &  1.6216 \\
$2/3$ & $\varepsilon^{MAP}$  & 1.3648   & 1.2426   & 1.2426 \\
\bottomrule
\end{tabular}
\label{table:r131223predictedthd}
\end{center}
\end{table}

\begin{figure*}[t]
    \centering
    \includegraphics[width=\linewidth]{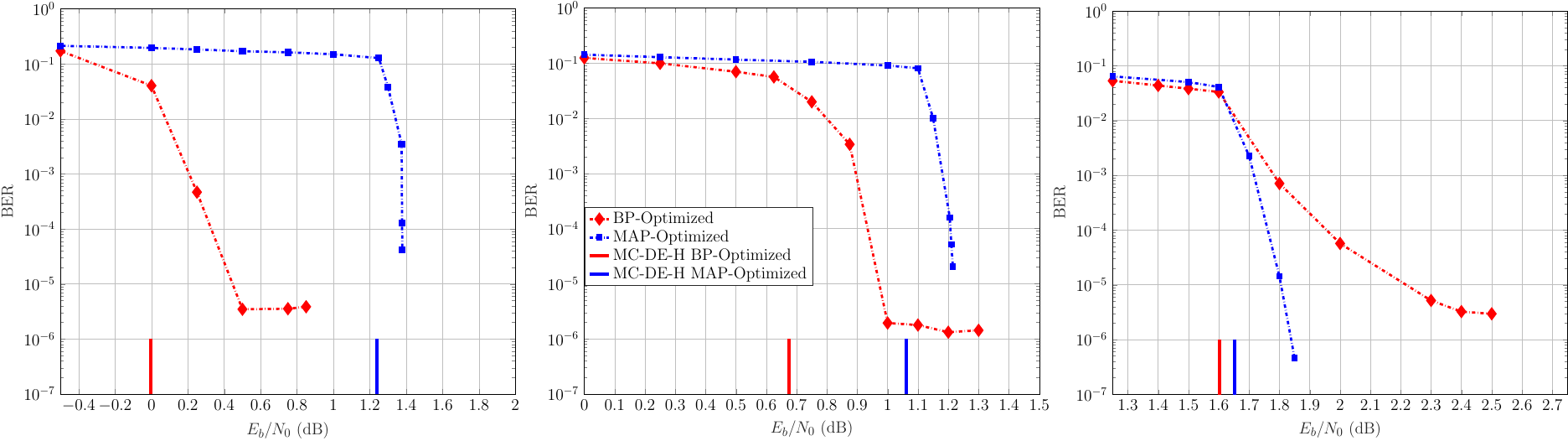}
    \caption{BER simulations for $n=102400$ and $R=1/3$ (left), $R=1/2$ (middle), and $R=2/3$ (right).}
    \label{fig:combinedLB} \medskip
\end{figure*}

BER simulations for randomly chosen CC-GLDPC codes of design rates $R=1/3,1/2,$ and $2/3$  with BP decoding %with the design configurations in Table~\ref{table:r131223bestBPMAP}
 are shown in Fig.~\ref{fig:combinedLB}. The maximum number of decoding iterations was set to 100 in all simulations. Each of the component code trellises was terminated separately, resulting in a small rate loss. Furthermore, the input block lengths were rounded to an integer resulting in a slightly different overall code rate.
All codes where constructed with a random permutation over the total number of edges $n_{e,1}+n_{e,1}=2n$, and also the positions of punctured bits in the mother code trellises were chosen randomly.
The AWGN channel thresholds shown in the figure correspond to Table~\ref{table:r13awgnthd}.

In general, the BER plots of the BP-optimized codes show a visible error floor and their gaps to the corresponding BP thresholds are larger than for MAP-optimized codes, indicating a slower convergence rate. The error floor is particularly bad for the highest code rate $R=2/3$. Note that for this ensemble a severe puncturing up to rate $r_2=0.9540$ is required.
Potentially, the error floors may be improved with structured ensembles and a more careful design of the permutations and the puncturing patters, but this is beyond the scope of this work.
The BER plots of the MAP-optimized codes, on the other hand, show steep decay in the waterfall regions and a lower gap to the corresponding BP thresholds. For lower code rates, however, they suffer from weak BP thresholds compared to the BP-optimized codes. This gap between the BP thresholds of MAP-optimized and BP-optimized codes gets smaller as the design rate $R$ increases, making MAP-optimized or even regular ensembles more attractive at high rates. At low rates, however, irregularity can significantly improve the performance in the waterfall region.

\begin{table}[t] 
\caption{MC-DE-H AWGN thresholds, $E_b/N_0(\rho^{*}) \, [\mathrm{dB}]$}
\begin{center}
\begin{tabular}{llrr}
\toprule
$R$ & Parameter &   Best BP & Best MAP \\
\midrule
$1/3$ & $(E_b/N_0)^{BP}$  & -0.0046   &  1.2386 \\
\midrule
$1/2$ & $(E_b/N_0)^{BP}$  & 0.6761   &     1.0616 \\
\midrule
$2/3$ & $(E_b/N_0)^{BP}$ & 1.6037 &    1.6516 \\
\bottomrule
\end{tabular}
\label{table:r13awgnthd}
\end{center}
\end{table}

\section{Concluding Remarks}
We have introduced different types of irregularity at the CNs of CC-GLDPC codes in order to improve their BP and MAP decoding thresholds. 
The proposed ensembles can be flexibly tuned in terms of trellis memory, fraction of edges connected to the different CNs, and the component code rates. 
An exhaustive grid search was then performed over these three dimensions to optimize the thresholds.
Our results show that irregularity can improve the BP thresholds compared to the regular case considered in \cite{muf_gldpctrellis} at the expense of a higher error floor.  

\newpage
The BP thresholds are also better than those of the turbo-like code ensembles considered in \cite{Moloudi_SCTC_Journal}, and we can avoid the use of degree-1 VNs.  
At larger rates, it can be observed that the gap between MAP-optimized and BP-optimized ensembles gets small and that irregularity becomes less advantageous.

\bibliographystyle{IEEEbib}
\bibliography{IEEEabrv,referencesJ}

 

% \end{IEEEbiography}



% \vfill

\end{document}